\documentclass[twocolumn,prl,superscriptaddress,showpacs]{revtex4}
\usepackage[utf8]{inputenc}
\usepackage{amsmath}
\usepackage{amssymb}
\usepackage{graphicx}
\usepackage{esint}
\usepackage{float}
\usepackage[unicode=true,pdfusetitle,
 bookmarks=true,bookmarksnumbered=false,bookmarksopen=false,
 breaklinks=false,pdfborder={0 0 1},backref=section,colorlinks=false]
 {hyperref}
\hypersetup{
 colorlinks,citecolor=blue}
\usepackage{breakurl}

\makeatletter
\@ifundefined{textcolor}{}
{%
 \definecolor{BLACK}{gray}{0}
 \definecolor{WHITE}{gray}{1}
 \definecolor{RED}{rgb}{1,0,0}
 \definecolor{GREEN}{rgb}{0,1,0}
 \definecolor{BLUE}{rgb}{0,0,1}
 \definecolor{CYAN}{cmyk}{1,0,0,0}
 \definecolor{MAGENTA}{cmyk}{0,1,0,0}
 \definecolor{YELLOW}{cmyk}{0,0,1,0}
}

\usepackage{subfigure}\usepackage{epsfig}\usepackage{amsfonts}\usepackage{mathrsfs}\usepackage[toc,page,title,titletoc,header]{appendix}
\usepackage{amsmath}

\makeatother

\begin{document}

\title{Center of Mass Momentum Dependent Interaction Between Ultracold Atoms}

\author{Jianwen Jie}

\affiliation{Department of Physics, Renmin University of China, Beijing, 100872,
China}

\author{Peng Zhang}
\email{pengzhang@ruc.edu.cn}


\affiliation{Department of Physics, Renmin University of China, Beijing, 100872,
China}

\affiliation{Beijing Computational Science Research Center, Beijing, 100084, China}

\affiliation{Beijing Key Laboratory of Opto-electronic Functional Materials \&
Micro-nano Devices (Renmin University of China)}

\begin{abstract}
We show that
a new type of two-body interaction, which
depends on the  momentum of the center of mass (CoM) of these two particles,
can be realized in ultracold atom gases with 
a laser-modulated magnetic Feshbach resonance (MFR). 
Here the MFR is modulated by two 
laser beams propagating along different directions, which
can  induce Raman transition between two-body bound
states. The Doppler effect causes the two-atom scattering length to be strongly dependent on the CoM momentum of these two atoms. 
As a result, the effective two-atom interaction is CoM-momentum dependent, while the one-atom free Hamiltonian is still the simple kinetic energy ${\bf p}^2/(2m)$. 
\end{abstract}


\maketitle

\emph{Introduction.}  In most physical systems the interaction between two particles
is a function of
their
relative position, and is independent of the two-body center of mass (CoM) 
degree of freedom. In some other systems, {\it e.g.} ultracold atom gases 
with interatomic scattering length being modulated by an inhomogeneous magnetic or laser field \cite{chinprl},
 the two-body interaction can be dependent on the CoM position. In this letter we show that  a new type of interaction, which depends on the two-body CoM {\it momentum},
 can be realized in ultracold atom gases, while the one-body free Hamiltonian remains the simple kinetic energy. Explicitly, we propose an approach to realizing an ultracold atom gas where the Hamiltonian of every two atoms can be formally expressed as

  \begin{eqnarray}
H_{\rm 2b}=\frac{{\bf p}_1^2}{2m_1}+\frac{{\bf p}_2^2}{2m_2}+V_{\rm eff}({\bf r}_1-{\bf r}_2, {\bf p}_1+{\bf p}_2),\label{h22b}
\end{eqnarray}
with $m_i$, ${\bf p}_i$ and ${\bf r}_i$ ($i=1,2$) being the mass, momentum and position of the $i$th atom, and $V_{\rm eff}$ being the CoM-momentum dependent effective two-atom interaction. To our knowledge, this type of interaction, which couples the two-body CoM motion and the relative motion without breaking the translational symmetry and changing the one-body dispersion relation, has not been discovered in any quantum system.
 
Our proposal is based on the magnetic Feshbach resonance (MFR)  modulated by two Raman laser beams  propagating along different directions,
which couple
the two-body bound states in the closed channel and are far off resonant for 
one-body transitions. When the interatomic interaction of an ultracold gas is controlled by such an approach, the two-atom scattering length is determined by the frequencies of these two laser beams. Furthermore, due to the Doppler effect, when the atoms are moving the frequency
of the laser beams can be effectively shifted, and the magnitude of the frequency shift 
depends on the two-atom CoM momentum. As a result, in this system the two-atom scattering length, which describes the effective two-atom interaction, becomes CoM-momentum dependent. In addition, the two-body collisional loss induced by spontaneous emission from excited state atoms can be significantly suppressed by the molecular dark state effect \cite{wuprl}.

 \begin{figure}
\includegraphics[width=7cm]{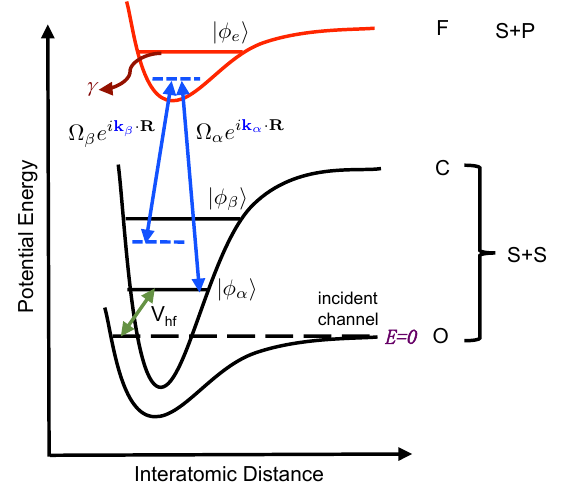} \caption{(color online) Schematic diagram for the MFR modulated by Raman laser
beams propagating along different directions (i.e., ${\bf k}_\alpha\neq{\bf k}_\beta$).
}
\end{figure}

In 
all the previous research for the optical control of interaction between ultracold atoms
\cite{OpticalFR,OpticalFRb,OpticalFR1,OpticalFR2,OpticalFR2b,OpticalFR3,OpticalFR4,bbtheory,bbexp,jingbb,wuprl,wupra,thomasprl,chinprl,socFR,socFR1,socFR2,socFR3,pengprl,hupra,qiprl}, the Doppler effect has always been ignored. This can be explained as follows.
In these control
processes
 the laser beams should be far off resonant to the two-body transitions
so that the collisional
loss induced by atomic spontaneous emission
can be suppressed. As a result, the Doppler shift of the
laser frequency is much smaller than the detuning of the laser-induced two-body transitions, and thus the Doppler effect
is negligible. However, in our system 
the two-atom scattering length depends on not only the
one-photon detuning but also the two-photon detuning of the laser-induced two-body Raman
transition. 
Since the two-photon detuning can be very
small, it can be significantly changed by the Doppler frequency shift
of the Raman laser beams.
Therefore,
the Doppler effect can be very important.

\textit{Three-dimensional (3D) $s$-wave scattering length.} We consider the $s$-wave
scattering of two ultracold alkali atoms in the ground electronic
orbital state (i.e., the S-state). As shown in
Fig.~1,
we assume
these two atoms are incident from the open channel $O$
 corresponding
to one specific two-atom hyperfine state. The threshold of this channel is
near resonant to a bound state $|\phi_{\alpha}\rangle$ in the closed channel
$C$, which corresponds to another hyperfine state of these two S-state atoms.
The energy difference between $|\phi_{\alpha}\rangle$ and the threshold
of channel $O$ can be controlled by the magnetic field. Thus,
a MFR \cite{MFR} can be induced by
the hyperfine coupling $V_{\rm hf}$ between the open channel $O$ and $|\phi_{\alpha}\rangle$.
We further
assume that a laser beam $\alpha$ with wave vector ${\bf k}_{\alpha}$
is applied to couple $|\phi_{\alpha}\rangle$ to another two-body bound state
$|\phi_{e}\rangle$ in an excited channel $F$ where one atom is in the S-state
and another atom is in the excited electronic orbital state (i.e.,
the P-state). The excited molecular state $|\phi_{e}\rangle$ can decay via the atomic spontaneous
emission process. In addition, $|\phi_{e}\rangle$ is also coupled
to a bound state $|\phi_{\beta}\rangle$ in the closed channel $C$, by another
laser beam $\beta$ with wave vector ${\bf k}_{\beta}$. As mentioned above, we assume the laser beams $\alpha$ and $\beta$ are
far off resonant for all the one-atom transition processes. As a result, they only induce two-body transitions and
do not change the one-atom Hamiltonian.

In our system the scattering length $a$ can be controlled by both
the magnetic field and the laser beams $\alpha$ and $\beta$. 
As we will show below,
when these two beams propagate
along the same direction, the Doppler effect is negligible and the scattering
length is still independent of the two-atom CoM momentum.
The control of the interatomic interaction for this case was proposed
by H. Wu and J. Thomas in Refs. \cite{wuprl,wupra} in 2011, where
the wave vectors ${\bf k}_{\alpha,\beta}$ were reasonably ignored.
It was experimentally realized by A. Jagannathan \textit{et al}.
in 2016 \cite{thomasprl}.

In this letter we consider the case where the two laser beams are
propagating along different directions, i.e. ${\bf k}_{\alpha}\neq{\bf k}_{\beta}$.
We will show that in this case the Doppler effect can be very important. As a result, the scattering length becomes significantly dependent
on the two-atom CoM momentum.

Now we calculate the scattering length $a$. Here we first
ignore the spontaneous decay of the excited molecular state $|\phi_{e}\rangle$ and illustrate the approach of our calculation. Then
we will take this decay into account and derive the explicit expression
for $a$.
In the absence of the spontaneous decay, in the Schr$\ddot{{\rm o}}$dinger
picture the two-body Hamiltonian can
be written as ($\hbar=1$) 
\begin{eqnarray}
H_{S} & = & \frac{{\bf P}^{2}}{2M}+H_{{\rm MFR}}+E_{\beta}|\phi_{\beta}\rangle\langle\phi_{\beta}|+E_{e}|\phi_{e}\rangle\langle\phi_{e}|,\nonumber \\
 &  & +\sum_{l=\alpha,\beta}\Omega_{l}e^{i({\bf k}_{l}\cdot{\bf R}-\omega_{l}t)}|\phi_{e}\rangle\langle\phi_{l}|+h.c.,\label{hs}
\end{eqnarray}
with $H_{{\rm MFR}}$ being defined as 
\begin{eqnarray}
H_{{\rm MFR}} & = & \left[\frac{{\bf p}^{2}}{2\mu}+V_{{\rm bg}}(r)\right]|O\rangle_{I}\langle O|+E_{\alpha}|\phi_{\alpha}\rangle\langle\phi_{\alpha}|\nonumber \\
 &  & +V_{\rm hf}(r)|O\rangle_{I}\langle C|+h.c..\label{hs0}
\end{eqnarray}
Here 
$|j\rangle_I$
($j=O,C,F$) denote the two-body internal state corresponding to channel $j$
\cite{symbol}, $M$ is the total mass and $\mu$ is the two-atom reduced mass,
 $E_{l}$ ($l=\alpha,e,\beta$) is
the energy of the bound state $|\phi_{l}\rangle$, $\omega_{l}$ ($l=\alpha,\beta$)
is the frequency of laser beam $l$, and $\Omega_{l}$ ($l=\alpha,\beta$)
is the strength of the laser-induced coupling between 
$|\phi_{l}\rangle$ and $|\phi_{e}\rangle$.
We have chosen the zero-energy point as the threshold of the open channel $O$.
 In Eqs. (\ref{hs}, \ref{hs0})
${\bf R}$ and ${\bf P}$ are the coordinate and momentum of the CoM,
while ${\bf r}$ and ${\bf p}$ are those of the
two-atom relative motion. The interaction in the open channel
$O$ is $V_{{\rm bg}}(r)$, and the hyperfine coupling between channels
$O$ and $C$ is described by $V_{\rm hf}(r)$.

We can simplify our problem and remove the phase factor $e^{\pm i({\bf k}_{l}\cdot{\bf R}-\omega_{l}t)}$
($l=\alpha,\beta$) by introducing a rotated
frame (interaction picture) where quantum state $|\Psi\rangle_{{\rm rot}}$
is related to the state $|\Psi\rangle_{S}$ in the Schr$\ddot{{\rm o}}$dinger
picture via the relation $|\Psi\rangle_{{\rm rot}}={\cal U}|\Psi\rangle_{S}$,
Here the unitary transformation ${\cal U}$ is given by 
\begin{eqnarray}
{\cal U} =  e^{i(\omega_{\alpha}t-{\bf k_{\alpha}}\cdot{\bf R})|\phi_{e}\rangle\langle\phi_{e}|}e^{i[(\omega_{\alpha}-\omega_{\beta})t-({\bf k_{\alpha}}-{\bf k}_{\beta})\cdot{\bf R}]|\phi_{\beta}\rangle\langle\phi_{\beta}|}.\label{u}
\end{eqnarray}
In this rotated frame, the two-body Hamiltonian becomes
\begin{eqnarray}
H_{{\rm rot}} & = & \frac{{\bf P}^{2}}{2M}+H_{{\rm MFR}}+\sum_{l=\alpha,\beta}\Omega_{l}|\phi_{e}\rangle\langle\phi_{l}|+h.c.\nonumber \\
 &  & +\Delta_{\rm 1p}({\bf P})|\phi_{e}\rangle\langle\phi_{e}|+\Delta_{\rm 2p}({\bf P})|\phi_{\beta}\rangle\langle\phi_{\beta}|,
\label{hrot}
\end{eqnarray}
where
\begin{eqnarray}
\Delta_{\rm 1p}({\bf P}) & = & \Delta_{\rm 1p}^{(0)}+\frac{|{\bf k}_{\alpha}|^{2}}{2M}+\frac{{\bf k}_{\alpha}\cdot{\bf P}}{M};\label{de}\\
\Delta_{\rm 2p}({\bf P}) & = & \Delta_{\rm 2p}^{(0)}+\frac{|{\bf k}_{\alpha}-{\bf k}_{\beta}|^{2}}{2M}+\frac{({\bf k}_{\alpha}-{\bf k}_{\beta})\cdot{\bf P}}{M},\label{db}
\end{eqnarray}
with $\Delta_{\rm 1p}^{(0)}=E_e-\omega_\alpha$ and $\Delta_{\rm 2p}^{(0)}=E_\beta-(\omega_\alpha-\omega_\beta)$.
The physical meaning of $\Delta_{\rm 1p}({\bf P})$ and $\Delta_{\rm 2p}({\bf P})$ can be understood as follows. Scattering length $a$ is determined 
by the multi-order transition process from the open channel $O$ to the bound state $|\phi_\alpha\rangle$ (induced by hyperfine coupling) and then to the excited molecular state $|\phi_e\rangle$ (induced by laser $\alpha$), and finally to the bound state $|\phi_\beta\rangle$ (induced by laser $\beta$). This is essentially a Raman process induced by the laser beams $\alpha$ and $\beta$. $\Delta_{\rm 1p}({\bf P})$ given by Eq. (\ref{de}) is the one-photon detuning of this Raman process (i.e., the detuning of the transition $O\rightarrow|\phi_\alpha\rangle\rightarrow|\phi_e\rangle$). Similarly, $\Delta_{\rm 2p}({\bf P})$ is the two-photon detuning of the complete Raman process from $O$ to $|\phi_\beta\rangle$.
Moreover, in Eqs. (\ref{de})
 and (\ref{db}) $\Delta_{\rm 1p}^{(0)}$ and $\Delta_{\rm 2p}^{(0)}$ 
can be understood as the bare value of these detuning, 
${|{\bf k}_{\alpha}|^{2}}/{(2M)}$
and ${|{\bf k}_{\alpha}-{\bf k}_{\beta}|^{2}}/{(2M)}$
are the shifts induced by the momentum-recoil effects, and the 
${\bf P}$-dependent terms are 
the Doppler shifts.

\begin{figure}[t]
\includegraphics[width=9cm]{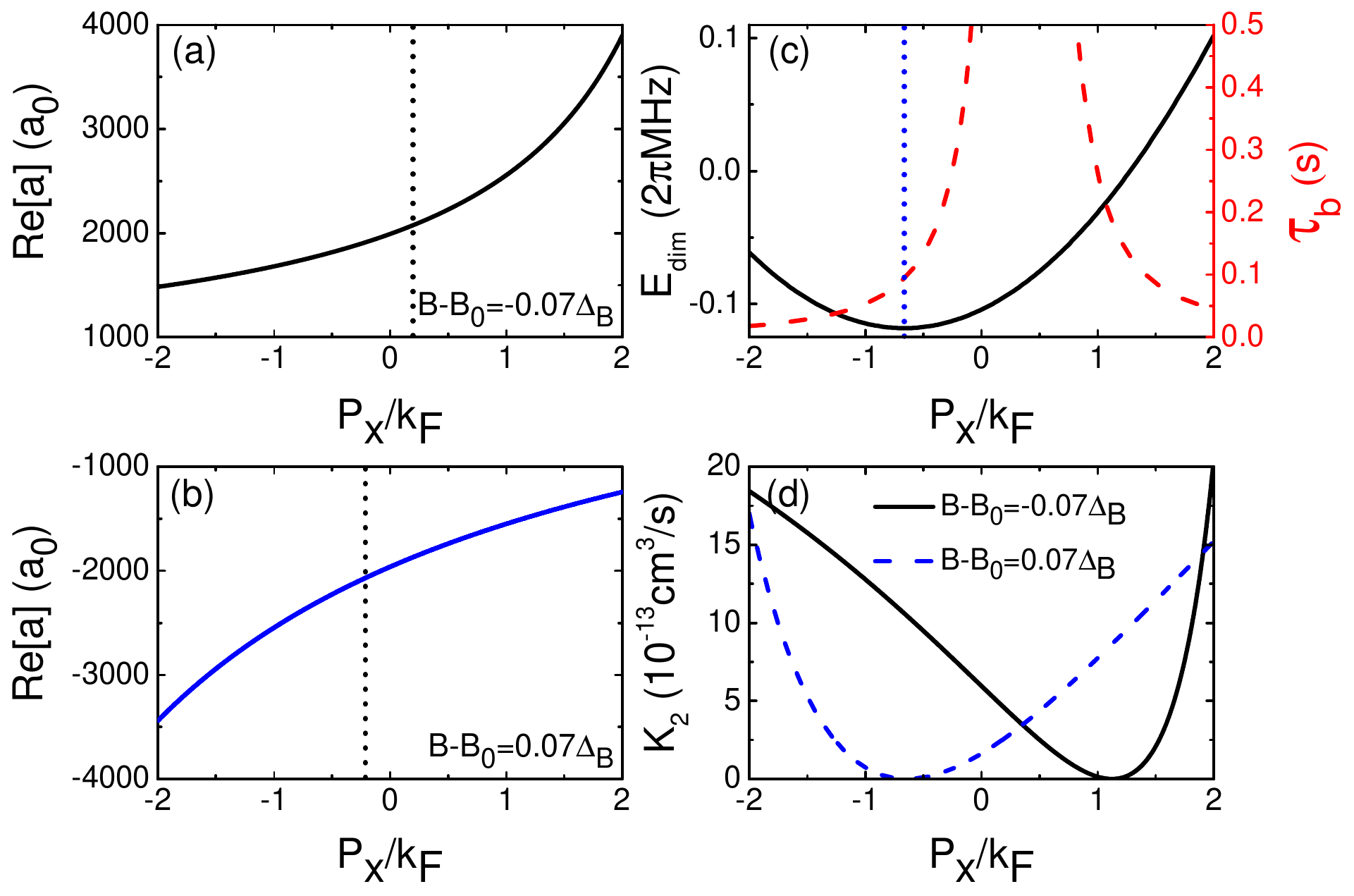} 
\caption{
(color online) 
{\bf (a)} and {\bf (b)}: ${\rm Re}[a(P_x)]$ of $^{40}$K
atoms with $B-B_{0}=\mp 0.07\Delta_{B}$. The black dotted line indicates the point  where $1/({k_{F}|{\rm Re}[a]|})=1$. 
{\bf (c)}: The dispersion relation 
$E_{\rm dim}(P_x)\equiv\frac{P_x^2}{4m}+{\rm Re}[E_b(P_x)]$ (black solid line) 
and lifetime $\tau_b(P_x)$ (red dashed line)  of the most shallow dimer for the case in (a). The blue dotted line indicates the point with minimum total dimer energy. 
{\bf (d)}: The 
loss rate $K_2(P_x)$ for the cases in (a, b).
In the calculations we take
$\Delta_{\rm 1p}^{(0)}+|{\bf k}_\alpha|^2/(2M)=2\pi\times400$MHz while $\Delta_{\rm 2p}^{(0)}+|{\bf k}_{\alpha}-{\bf k}_{\beta}|^{2}/(2M)=-2\pi\times2.1\times10^{4}$Hz for
(a) and $2\pi\times1.2\times10^{4}$Hz for (b). 
In the experiment of laser-modulated MFR of $^{40}$K atoms \cite{jingbb}, the laser-induced
coupling intensity between bound states can be as large as $30$MHz. Thus, here we choose
$\Omega_{\alpha}=2\pi\times50$MHz while
$\Omega_{\beta}=2\pi\times10$MHz for (a)
and $2\pi\times12$MHz for (b).
We further choose
$\omega_{a}\approx\omega_{b}=2\pi\times3.9\times10^{14}$Hz and $\gamma=2\pi\times6$MHz,
respectively \cite{jingbb}. Other parameters are given in the main text.
}
\end{figure}

According to Eq. (\ref{u}), we have ${\cal U}|O\rangle_{I}\langle O|{\cal U}^{\dagger}=|O\rangle_{I}\langle O|$. Thus, all the operators for the open channel $O$ are 
unchanged under the frame transformation.
Therefore, we can calculate the scattering length of two atoms incident
from channel $O$, which is determined by the behavior of the low-energy
scattering wave function in this channel in the limit $r\rightarrow\infty$,
by solving the scattering problem in the rotated frame, which is governed by $H_{\rm rot}$.
Furthermore,
Eq. (\ref{hrot}) shows that in this frame
the CoM momentum ${\bf P}$ is conserved.
Thus, the scattering
length $a$ is a function of ${\bf P}$.

Now we consider the spontaneous decay of the excited molecular state $|\phi_{e}\rangle$.
We can theoretically take into account this decay
by introducing an auxiliary scattering channel which is coupled
to $|\phi_{e}\rangle$ \cite{paulOpticalFR}. With this approach we derive the exact analytical expression of the scattering
length \cite{supp}: 
\begin{eqnarray}
a({\bf P})\!\!&=&\!\!a_{{\rm bg}}\!-\!\frac{(\delta\mu)a_{{\rm bg}}\Delta_{B}\!\left[\Delta_{\rm 1p}({\bf P})-i\frac{\gamma}{2}-\frac{|\Omega_{\beta}|^{2}}{{\Delta}_{\rm 2p}({\bf P})}\right]}{(\delta\mu)(B\!-\!B_{0})\!\left[{\Delta}_{\rm 1p}({\bf P})-i\frac{\gamma}{2}-\frac{|\Omega_{\beta}|^{2}}{{\Delta}_{\rm 2p}({\bf P})}\right]\!-|\Omega_{\alpha}|^{2}}.\nonumber\\
\ \label{ap}
\end{eqnarray}
Here
 $a_{{\rm bg}}$, $B_{0}$ and $\Delta_{B}$ are the background
scattering length, the resonance position and the width of the MFR. Explicitly,
in the absence of the Raman laser beams we have $a=a_{{\rm bg}}[1-\Delta_{B}/(B-B_{0})]$.
$\delta\mu$ is the magnetic moment difference between the channels
$O$ and $C$, and $\gamma$ is the spontaneous
decay rate of $|\phi_{e}\rangle$. The 
energy of two-body bound state can also be calculated via the same approach.

It is clear that the scattering length $a({\bf P})$ depends on the
CoM momentum ${\bf P}$ via 
the one-photon and two-photon detuning
${\Delta}_{\rm 1p}({\bf P})$
and ${\Delta}_{\rm 2p}({\bf P})$. 
In a realistic system, to suppress the collisional loss induced by the spontaneous decay of $|\phi_e\rangle$, one usually sets the bare value
$\Delta_{\rm 1p}^{(0)}$
 of the one-photon detuning 
  to be very large. 
As a result, the ${\bf P}$-dependence of  ${\Delta}_{\rm 1p}({\bf P})$ is usually negligible. However,
the bare two-photon detuning $\Delta_{\rm 2p}^{(0)}$ can be very
small, and thus the variation of ${\Delta}_{\rm 2p}({\bf P})$ with ${\bf P}$ can be very
significant. Therefore, $a({\bf P})$ can be
strongly dependent on ${\bf P}$ via ${\Delta}_{\rm 2p}({\bf P})$.

Furthermore, according to Eq. (\ref{db}), when the two
Raman beams are propagating along the same direction (i.e., ${\bf k}_\alpha\approx{\bf k}_\beta$), ${\Delta}_{\rm 2p}({\bf P})$ takes a ${\bf P}$-independent value $\Delta_{\rm 2p}^{(0)}$, and thus the Doppler effect can be ignored.
For this case
it was shown that the two-body loss can be suppressed by the molecular dark state effect, provided that
${\Delta}_{\rm 2p}^{(0)}$ is small enough \cite{wuprl}. 
Since
we can re-obtain the scattering length for this case \cite{wuprl} by replacing ${\Delta}_{\rm 2p}({\bf P})$ in Eq. (\ref{ap}) with ${\Delta}_{\rm 2p}^{(0)}$, 
for our system
the two-body loss can also be suppressed when ${\Delta}_{\rm 2p}({\bf P})$
is small enough. This suppression can also be understood from the fact $\lim_{{\Delta}_{\rm 2p}({\bf P})\rightarrow0}{\rm Im}[a({\bf P})]\propto{\cal O}({\Delta}_{\rm 2p}({\bf P})^2)$.

\textit{3D ultracold Fermi gas.} As an example, we consider
a 3D ultracold gas of two-component $^{40}$K atoms in the lowest
two hyperfine states $|\uparrow\rangle\equiv|F=9/2,m_{F}=-9/2\rangle$
and $|\downarrow\rangle\equiv|F=9/2,m_{F}=-7/2\rangle$. 
Here we focus on the MFR with $B_{0}=202.2$ G,
$\Delta_{B}=8$ G and $a_{{\rm bg}}=174a_{0}$ \cite{jingbb,MFR}, with $a_{0}$
being the Bohr radius,
and assume
that this MFR modulated by two Raman
beams as discussed above. 
We take these two beams to be counter-propagating along the $x$-axis.
Thus, the scattering length $a$ only depends on
the $x$-component $P_x$ of ${\bf P}$. In the ultracold Fermi gas 
$P_{x}$ is mainly in the region between $-2k_{F}$ and $2k_{F}$,
with $k_{F}$ being the Fermi momentum. 
We consider an ultracold gas with Fermi temperature $T_{F}=0.5\mu$K (corresponding to $k_F=9.1\times10^6{\rm  m}^{-1}$). 

In Fig.~2(a, b) we illustrate the
real part of $a$ given by Eq. (\ref{ap})
for $B-B_{0}=\mp 0.07\Delta_{B}$.
It is shown that in these cases ${\rm Re}[a]$ is always
positive or always negative for $P_{x}\in[-2k_{F},2k_{F}]$, and can
change by about $2500 a_{0}$ with $P_{x}$, ranging from the region with $1/(k_F|{\rm Re}[a]|)<1$ to the region with $1/(k_F|{\rm Re}[a]|)>1$. Direct calculations show that these results are robust with respect to the uncertainties of $B_0$ and $\Delta_B$.
For the case of Fig.~2(a) we further calculate the energy $E_b$ of the most shallow two-body bound state, as function of the CoM momentum $P_x$. In Fig.~2(c) we show the total dimer energy $E_{\rm dim}(P_x)\equiv\frac{P_x^2}{2M}+{\rm Re}[E_b(P_x)]$ as a function of $P_x$, i.e. the dispersion relation of the shallow dimer. It is clear that if the $E_b$ were independent of $P_x$, the minimum point of $E_{\rm dim}$ appears at $P_x=0$.
As shown in Fig.~2(c), due to the $P_x$-dependence of $E_b$, in our system
$E(P_x)$ takes its minimum value when 
$P_x=-0.66k_F$.


Now we investigate the two-atom collisional loss. If the scattering
length $a$ was independent of the CoM momentum, 
the two-body collisional loss rate is $K_{2}\equiv-8\pi\hbar\text{Im}[a]/m$, with 
$m$ being the single-atom mass. 
Accordingly, the lifetime of the ultracold
gas can be defined as $\tau=1/[K_{2}n_0]$, where $n_0$ is the initial atom density.
When $a$
is $P_{x}$-dependent, it is difficult to exactly calculate
the two-body loss rate and lifetime for the gas. Nevertheless,
we can still estimate the loss effect via the
parameter $K_{2}(P_{x})\equiv-8\pi\hbar\text{Im}[a(P_{x})]/m$.
As shown in
Fig.~2(d),  $K_{2}(P_{x})$ is
below $2\times10^{-12}{\rm cm}^{3}/{\rm s}$ for the systems studied in Fig.2~(a, b).
Using the atom density $n_0=1.28\times 10^{13}/{\rm cm}^3$
corresponding to $T_F=0.5 \mu$K, we can obtain
$1/[K_{2}(P_{x})n_0]>0.04$s. 
In addition, in  Fig.~2(d) we illustrate the lifetime $\tau_b\equiv1/({\rm Im}[E_b(P_x)])$
of the shallow dimer. It is shown that $\tau_b$ is about $0.1$s 
at the minimum point of the dimer energy.

\textit{Quasi one dimensional (quasi-1D) ultracold Fermi gas.} 
Now we consider
an ultracold two-component Fermi gas in an axially symmetric two-dimensional harmonic
potential in the $y-z$ plane, with trapping frequency $\omega_{\perp}$.
When the atomic transverse motion is frozen in the ground state of
this harmonic potential, the ultracold gas becomes a quasi-1D system. For this system the effective low-energy 1D interaction between two
atoms in different components can be expressed as 
\begin{eqnarray}
V_{{\rm 1D}}=g_{{\rm 1D}}\delta(x)\equiv-\frac{1}{\mu a_{{\rm 1D}}}\delta(x).\label{g1d}
\end{eqnarray}
Here $a_{{\rm 1D}}$ is
 the effective 1D scattering length. It can be controlled by
 both the 3D scattering length and the transverse trapping frequency $\omega_{\perp}$ via the confinement-induced resonance (CIR) effect
  \cite{olshanii}.
As shown above, when
the two-atom interaction is controlled by a MFR modulated by two Raman
beams counter-propagating along the $x$-axis, the 3D scattering length becomes a function
of the CoM momentum $P_{x}$.
As a result, both $a_{{\rm 1D}}$ and the effective 1D interaction intensity
$g_{{\rm 1D}}$ become $P_{x}$-dependent.

In particular, 
when $a_{\rm 1D}(P_x=0)$ is tuned to the CIR point, i.e. when $a_{\rm 1D}(P_x=0)=0$, one
can even obtain a dramatic quasi-1D system with 
\begin{equation}
g_{{\rm 1D}}(P_{x})=\left\{
\begin{array}{rcl}
<0       &      & {{\rm for}\ P_x>0}\\
=\infty     &      & {{\rm for}\ P_x=0}\\
>0     &      & {{\rm for}\ P_x<0}
\end{array} \right..
\end{equation}
Namely, in this system the atoms have attractive and repulsive 1D
interactions when the CoM momentum is along the $+x$ and $-x$ direction, respectively,
and have an infinitely strong interaction when the CoM momentum
is zero. 
In Fig.~3  we illustrate such a case for a quasi-1D ultracold gas of $^{40}$K
atoms in hyperfine states $|\uparrow\rangle$ and $|\downarrow\rangle$.
The lifetime of this quasi-1D gas is estimated to be larger than $0.04$s \cite{supp}.
.

\begin{figure}[t]
\includegraphics[width=5.5cm]{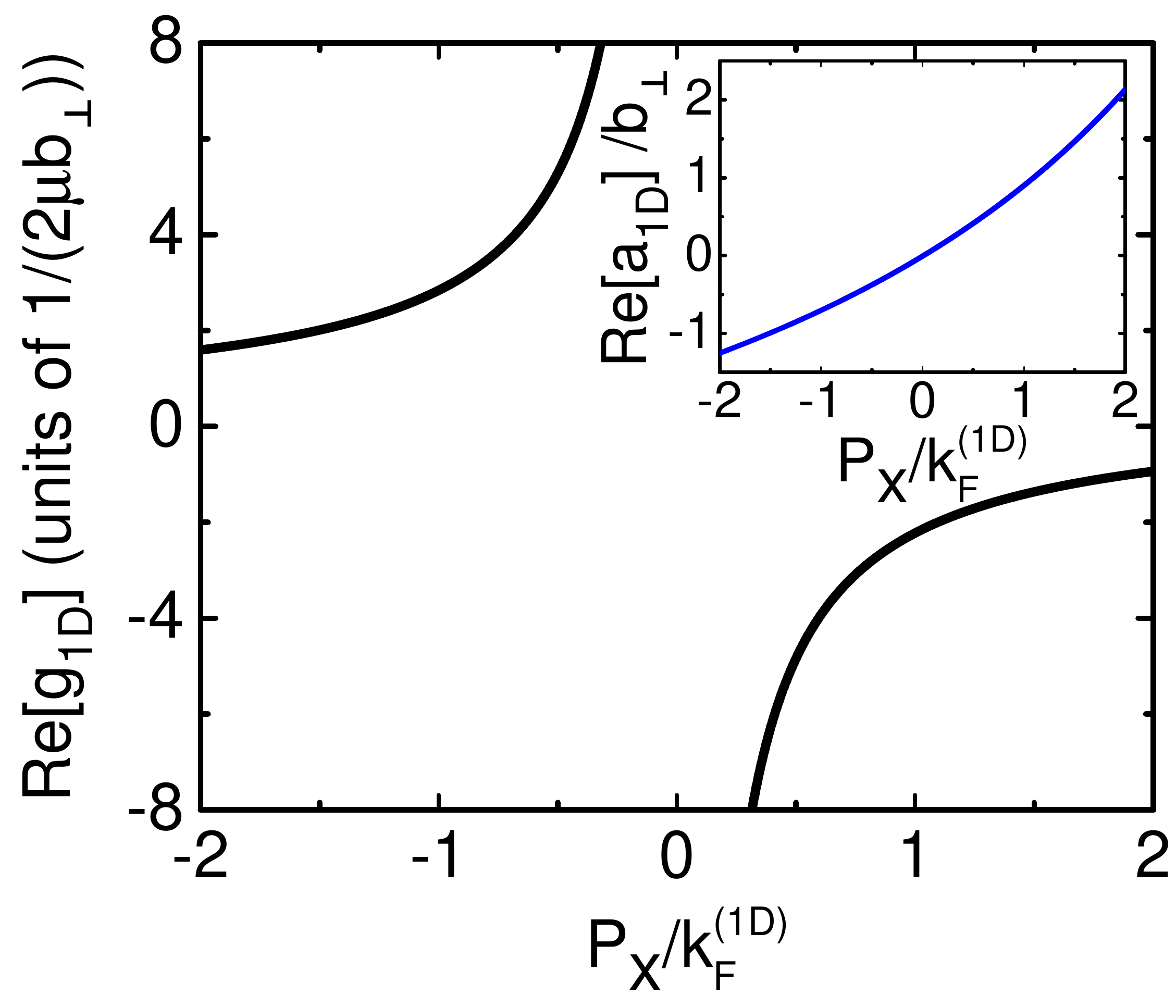} 
\caption{(color online)  
${\rm Re}[g_{{\rm 1D}}(P_{x})]$ and ${\rm Re}[a_{{\rm 1D}}(P_{x})]$
(inset) of the quasi-1D ultracold gas of $^{40}$K atoms. 
We consider the system where the one-body momentum $k_x$ along the $x$-direction is in the region between $\pm k_F^{\rm (1D)}\equiv\pm3/(4b_{\perp})$ with $b_\perp=\sqrt{2/(m\omega_\perp)}$. Thus we have $P_x\in[-2 k_F^{\rm (1D)}, 2k_F^{\rm (1D)}]$.
In our calculation
we take $\Delta_{\rm 1p}^{(0)}+|{\bf k}_\alpha|^2/(2M)=2\pi\times100$MHz, 
$\Delta_{\rm 2p}^{(0)}+|{\bf k}_{\alpha}-{\bf k}_{\beta}|^{2}/(2M)=2\pi\times1.1\times10^{4}$Hz,
$\Omega_{\alpha}=2\pi\times50$MHz, $\Omega_{\beta}=2\pi\times3.5$MHz, $b_{\perp}=4000a_{0}$
 and 
$B-B_{0}=-0.07\Delta_{B}$. Other parameters
are the same as in Fig.~2.}
\end{figure}


\textit{Summary and discussion.} 
In common quantum systems, the two-body CoM and relative motion can be coupled by
a non-harmonic confinement potential, a spin-orbit coupling or a CoM-position-dependent two-body interaction. Nevertheless, these approaches either destroy the translational symmetry or change the one-body dispersion relation. Here we show that a CoM-momentum dependent interaction between 
ultracold atoms can be realized via
a laser-modulated MFR. This interaction can couple the CoM and relative motion without
breaking translational symmetry or changing the
one-body dispersion relation.
Thus, various new effects can be induced by this interaction. For instance, 
for a 3D two-component Fermi gas with CoM-momentum dependent positive scattering length $a({\bf P})$,  the dimers are possible to condense
in a superfluid state with nonzero momentum.
When $a({\bf P})$ is negative, it is possible that the minimum energy of a Cooper pair appears in the region with nonzero CoM momentum, and thus the Fulde-Ferrell phase \cite{FF} can be induced. 
In addition, the single-atom momentum distribution or contact relation can also be qualitatively modified by a CoM-momentum dependent interaction \cite{cui}.
Furthermore,  this type of interaction
can be used to study the 1D anyon model \cite{guan} or other
high-order nonlinear Schr$\ddot{\rm o}$dinger models
 \cite{q1da,q1db,q1dc}.

\textit{Acknowledgement.} We thank Paul Julienne, Haibin Wu and Hui Zhai for
helpful discussions. We also thank the referees for improving the 
quality of this manuscript.
This work has been supported by the Natural Science
Foundation of China under Grant No. 11434011, 11674393, and by NKBRSF of China
under Grant No. 2012CB922104, the Fundamental Research Funds for the
Central Universities, and the Research Funds of Renmin University
of China under Grant No.16XNLQ03, 17XNH054.


\begin{thebibliography}{10}
\bibitem{symbol}In our two-body problem the Hilbert space is ${\cal H}_{{\rm CoM}}\bigotimes{\cal H}_{{\rm rel}}\bigotimes{\cal H}_{{\rm internal}}$,
where ${\cal H}_{{\rm internal}}$ being the space for the two-atom
internal state, while ${\cal H}_{{\rm CoM}}$ and ${\cal H}_{{\rm rel}}$
are the spaces for the spatial states of CoM motion and relative motion,
respectively. Here we use $|\rangle_{I}$ to denote
the states in ${\cal H}_{{\rm internal}}$ and $|\rangle$  for those in ${\cal H}_{{\rm rel}}\bigotimes{\cal H}_{{\rm internal}}$.
\end{thebibliography}
\end{document}


\begin{widetext}

\section*{SUPPLEMENTARY MATERIAL}

In this supplementary material we will first prove Eqs. (5-7) in our main text, and then calculate the three-dimensional
(3D) scattering length for our system, with a model in which the spontaneous decay of the excited molecule state is included. We will further calculate
the energy and lifetime of the most
shallow two-body bound state for our system,
and discuss the robustness of our results with respect to the uncertainty of the magnetic Feshbach resonance (MFR) point and width. In the end of this  supplementary material we calculate the effective one-dimensional (1D) scattering
length and interaction intensity.

\section{Proof of Eqs. (5-7)}

As shown in our main text, the rotated frame introduced in our problem is defined
via the relation
\begin{equation}
  |\Psi\rangle_{{\rm rot}}={\cal U}|\Psi\rangle_{S}, \label{ee1},\tag{S1}
\end{equation}
where $|\Psi\rangle_{S}$ and $|\Psi\rangle_{{\rm rot}}$ are the states in the Schr$\ddot{{\rm o}}$dinger picture and the rotated frame, respectively, and ${\cal U}$ is the unitary transformation given by Eq. (4) of the main text. Substituting the relation (\ref{ee1}) into the Schr$\ddot{{\rm o}}$dinger equation
\begin{equation}
i\frac{d}{dt}|\Psi\rangle_{S}=H_S|\Psi\rangle_{S}, \label{e2}\tag{S2}
\end{equation}
with $H_S$ being given by Eq. (2) of the main text, we obtain the Schr$\ddot{{\rm o}}$dinger equation satisfied by the state in the rotated frame:
\begin{equation}
i\frac{d}{dt}|\Psi\rangle_{{\rm rot}}=H_{\rm rot}|\Psi\rangle_{\rm rot}, \label{e3}\tag{S3}
\end{equation}
where $H_{\rm rot}$ is the Hamiltonian in the rotated frame, and is given by
\begin{equation}
H_{\rm rot}={\cal U}H_S{\cal U}^\dagger+i\left(\frac{d}{dt}{\cal U}\right){\cal U}^\dagger.\label{e4}\tag{S4}
\end{equation}
Substituting Eq. (2) and (4) of our main text into Eq. (\ref{e4}), we can obtain Eqs. (5-7) of the main text. Here we have used the relations
\begin{align}
e^{i{\bf k}\cdot{\bf R}|\phi_j\rangle\langle\phi_j|}{\bf P}e^{-i{\bf k}\cdot{\bf R}|\phi_j\rangle\langle\phi_j|}&={\bf P}+{\bf k}|\phi_j\rangle\langle\phi_j|;\tag{S5}\\
e^{i{\bf k}\cdot{\bf R}|\phi_j\rangle\langle\phi_j|}\left({\bf P}^2\right)e^{-i{\bf k}\cdot{\bf R}|\phi_j\rangle\langle\phi_j|}&=\left({\bf P}+{\bf k}|\phi_j\rangle\langle\phi_j|\right)^2={\bf P}^2+\left({\bf k}^2+2{\bf k}\cdot{\bf P}\right)|\phi_j\rangle\langle\phi_j|\tag{S6}
\end{align}
for $j=e,\beta$, where ${\bf k}$ is a constant vector (c-number) while ${\bf P}$ and ${\bf R}$ are the {\it operators} for CoM momentum and position, respectively. They satisfy $[P_i,R_j]=i\hbar\delta_{i,j}$ for $i,j=x,y,z$, with $P_i$ and $R_j$ being the components of ${\bf P}$ and ${\bf R}$ in the directions $i$ and $j$.

\section{Explicit Expression for 3D Scattering Length}

Now we derive the explicit expression for 3D scattering length, i.e., Eq. (8) of our main text, with a model where the
spontaneous decay of the excited molecule state is taken into account.

\subsection{Model and Rotated Frame}

As shown in Ref. {[}23{]} of the main text, in our problem the Hilbert
space is ${\cal H}_{{\rm CoM}}\bigotimes{\cal H}_{{\rm rel}}\bigotimes{\cal H}_{{\rm internal}}$,
where ${\cal H}_{{\rm internal}}$ being the space for the two-atom
internal state, while ${\cal H}_{{\rm CoM}}$ and ${\cal H}_{{\rm rel}}$
are the spaces for the spatial states of CoM motion and relative motion,
respectively. Here we use $|\rangle_{I}$ and $|\rangle_{r}$ to denote
the states in ${\cal H}_{{\rm internal}}$ and ${\cal H}_{{\rm rel}}$
respectively, and use $|\rangle$ to denote the states in ${\cal H}_{{\rm rel}}\bigotimes{\cal H}_{{\rm internal}}$.

\begin{figure*}[t]
\includegraphics[width=9.5cm]{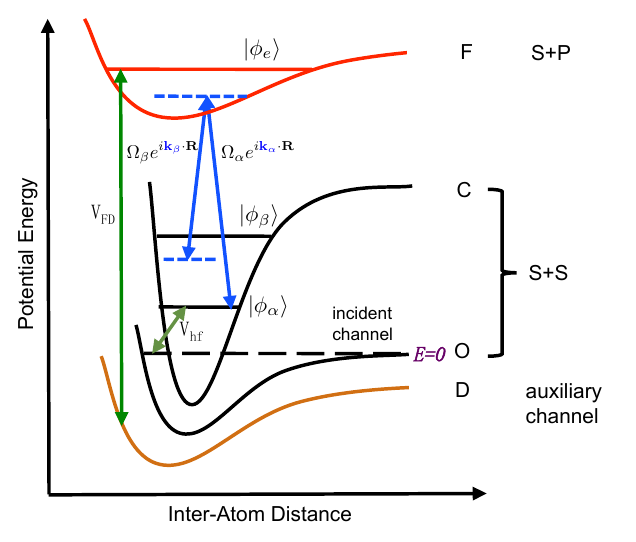}
\begin{center}
{FIG. S1: (color online) The model used in our calculation. }
\end{center}
\end{figure*}

As mentioned in the main text, we can theoretically take into account
the spontaneous decay of the excited molecular state $|\phi_{e}\rangle$
by introducing an auxiliary scattering channel $D$ which is coupled
to the channel $F$ (Fig.~S1) \cite{paulOpticalFRs}. The two-atom
internal state corresponding to this channel can be formally denoted
as $|D\rangle_{I}$. Accordingly, in the Schr$\ddot{{\rm o}}$dinger
picture the total Hamiltonian for the two atoms in the 3D space is
given by ($\hbar=1$)
\begin{equation}
H_{S}^{({\rm tot)}}=H_{S}+\left[\frac{{\bf p}^{2}}{2\mu}+E_{D}+V_{D}(r)\right]|D\rangle_{I}\langle D|+V_{FD}(r)|F\rangle_{I}\langle D|+h.c..,\tag{S7}
\end{equation}
where $H_{S}$ is defined in Eq. (2) of the main text, $E_{D}$
and $V_{D}(r)$ are the threshold energy and potential for the auxiliary
channel $D$, respectively, and $V_{FD}(r)$ describes the coupling
between channel $D$ and the excited channel $F$. Other notations
are defined in the main text. The energy gap $E_{e}-E_{D}$ between
the excited molecular state and the threshold of the auxiliary channel
$D$ is on the order of optical transition (i.e.,$\sim2\pi\times10^{14}$Hz).

As shown in the main text, we can simplify our problem and remove
the phase factor $e^{\pm i({\bf k}_{l}\cdot{\bf R}-\omega_{l}t)}$
($l=\alpha,\beta$) in the Hamiltonian by introducing the rotated
frame (i.e., the interaction picture). In the presence of the auxiliary
channel $D$, quantum state $|\Psi\rangle_{{\rm rot}}$ is related
to the state $|\Psi\rangle_{S}$ in the Schr$\ddot{{\rm o}}$dinger
picture via the relation $|\Psi\rangle_{{\rm rot}}=U|\Psi\rangle_{S}$,
with the unitary transformation $U$ being given by
\begin{equation}
U={\cal U}e^{i(\omega_{\alpha}t-{\bf k}_{\alpha}\cdot{\bf R})|D\rangle_{I}\langle D|}=e^{i(\omega_{\alpha}t-{\bf k_{\alpha}}\cdot{\bf R})(|\phi_{e}\rangle\langle\phi_{e}|+|D\rangle_{I}\langle D|)}e^{i[(\omega_{\alpha}-\omega_{\beta})t-({\bf k_{\alpha}}-{\bf k}_{\beta})\cdot{\bf R}]|\phi_{\beta}\rangle\langle\phi_{\beta}|},\tag{S8}
\end{equation}
where ${\cal U}$ is defined in Eq. (4) of the main text. In the rotated
frame, the total Hamiltonian is given by
\begin{equation}
H_{{\rm rot}}^{({\rm tot)}}=\frac{{\bf P}^{2}}{2M}+H_{{\rm rel}}({\bf P}).\tag{S9}
\end{equation}
Here $H_{{\rm rel}}({\bf P})$ is defined as
\begin{align}
H_{{\rm rel}}({\bf P}) & =H_{F}({\bf P})+Y+Z,\tag{S10}
\end{align}
with
\begin{align}
H_{F}({\bf P}) & =\left[\frac{{\bf p}^{2}}{2\mu}+V_{{\rm bg}}(r)\right]|O\rangle_{I}\langle O|+\left[\frac{{\bf p}^{2}}{2\mu}+\Delta_{{\rm 1p}}({\bf P})+E_{D}-E_{e}+V_{D}(r)\right]|D\rangle_{I}\langle D|\nonumber \\
 & +E_{\alpha}|\phi_{\alpha}\rangle\langle\phi_{\alpha}|+\Delta_{{\rm 1p}}({\bf P})|\phi_{e}\rangle\langle\phi_{e}|+\Delta_{{\rm 2p}}({\bf P})|\phi_{\beta}\rangle\langle\phi_{\beta}|,\tag{S11}
 \label{HF}
\end{align}
and
\begin{align}
Y & =V_{{\rm hf}}(r)|O\rangle_{I}\langle C|+V_{FD}(r)|F\rangle_{I}\langle D|+h.c.;\tag{S12}\\
Z & =\sum_{l=\alpha,\beta}\Omega_{l}|\phi_{e}\rangle\langle\phi_{l}|+h.c..\tag{S13}\label{S6}
\end{align}
In Eq. (\ref{HF}) the one-photon detuning $\Delta_{{\rm 1p}}({\bf P})$ and two-photon
detuning $\Delta_{{\rm 2p}}({\bf P})$ and other notations are all
defined in the main text.

It is clear that for a given CoM momentum ${\bf P}$, the two-body
relative motion is governed by $H_{{\rm rel}}({\bf P})$. In the following
we calculate the scattering length and two-body bound state energy
by solving the two-body problem with Hamiltonian $H_{{\rm rel}}({\bf P})$.

\subsection{3D Scattering State}

To calculate the scattering length and bound-state energy, here we
first calculate the 3D scattering state and scattering amplitude of
two atoms incident from the open channel $O$ with incident momentum
${\bf k}$. To this end we first derive the out-going scattering state
$|\Psi_{{\bf k}}^{(+)}\rangle$ for these two atoms which can be expanded
as
\begin{align}
|\Psi_{{\bf k}}^{(+)}\rangle & =|\psi_{{\bf k}}^{(O)}\rangle_{r}|O\rangle_{I}+|\psi_{{\bf k}}^{(D)}\rangle_{r}|D\rangle_{I}+\sum_{l=\alpha,e,\beta}b_{{\bf k}}^{(l)}|\phi_{l}\rangle\nonumber\\
&\equiv|\Phi_{{\bf k}}\rangle+\sum_{l=\alpha,e,\beta}b_{{\bf k}}^{(l)}|\phi_{l}\rangle
\tag{S14}\label{ss}
\end{align}
and satisfies the equation \cite{books}
\begin{align}
|\Psi_{{\bf k}}^{(+)}\rangle & =\lim_{\varepsilon\rightarrow0^{+}}\frac{i\varepsilon}{E_{{\bf k}}+i\varepsilon-H_{{\rm rel}}({\bf P})}|{\bf k}\rangle_{r}|O\rangle_{I}.\tag{S15}\label{lse1}
\end{align}
Here $E_{{\bf k}}={\bf k}^{2}/(2\mu)$ is the scattering energy, $|{\bf k}\rangle_{r}$
is the eigen state of the relative momentum operator ${\bf p}$, and
thus $|{\bf k}\rangle_{r}|O\rangle_{I}$ is the incident state of
these two atoms. Using the relation
\begin{align}
\frac{1}{E_{{\bf k}}+i\varepsilon-H_{{\rm rel}}({\bf P})} & =\frac{1}{E_{{\bf k}}+i\varepsilon-H_{F}}+\frac{1}{E_{{\bf k}}+i\varepsilon-H_{F}}\left(Y+Z\right)\frac{1}{E_{{\bf k}}+i\varepsilon-H_{{\rm rel}}({\bf P})},\tag{S16}
\end{align}
we can re-write Eq. (\ref{lse1}) as \cite{paulrmps}
\begin{align}
|\Psi_{{\bf k}}^{(+)}\rangle & =|\psi_{{\bf k}}^{{\rm bg(+)}}\rangle_{r}|O\rangle_{I}+G_{F}(E_{{\bf k}})(Y+Z)|\Psi_{{\bf k}}^{(+)}\rangle,\tag{S17}\label{lse2}
\end{align}
with
\begin{align}
|\psi_{{\bf k}}^{{\rm bg(+)}}\rangle_{r} & =\lim_{\varepsilon\rightarrow0^{+}}\frac{i\varepsilon}{E_{{\bf k}}+i\varepsilon-\left(\frac{{\bf p}^{2}}{2\mu}+V_{{\rm bg}}(r)\right)}|{\bf k}\rangle_{r}.\tag{S18}
\end{align}
being the outgoing scattering state for the open channel $O$ itself
and
\begin{align}
G_{F}(E)&=\frac{1}{E+i0^{+}-H_{F}}.\tag{S19}\label{phik}
\end{align}
Substituting Eq. (\ref{ss}) into Eq. (\ref{lse2}), we derive the
equaitons for $|\Phi_{{\bf k}}\rangle$ and the coefficients $b_{{\bf k}}^{(l)}$
($l=\alpha,e,\beta$):
\begin{align}
|\Phi_{{\bf k}}\rangle & =|\psi_{{\bf k}}^{{\rm bg(+)}}\rangle_{r}|O\rangle_{I}+G_{F}(E_{{\bf k}})Y\left[\sum_{l^{\prime}=\alpha,e,\beta}b_{{\bf k}}^{(l^{\prime})}|\phi_{l^{\prime}}\rangle\right];\tag{S20}\label{e1}\\
b_{{\bf k}}^{(l)} & =\frac{1}{E_{{\bf k}}-\Lambda_{l}({\bf P})}\left(\langle\phi_{l}|Y|\Phi_{{\bf k}}\rangle+\sum_{l^{\prime}=\alpha,e,\beta}\langle\phi_{l}|Z|\phi_{l^{\prime}}\rangle b_{{\bf k}}^{(l^{\prime})}\right)\ \ \ {\rm for}\ l=\alpha,e,\beta,\tag{S21}\label{e2}
\end{align}
with $\Lambda_{\alpha}({\bf P})=E_{\alpha}$, $\Lambda_{e}({\bf P})=\Delta_{{\rm 1p}}({\bf P})$ and $\Lambda_{\beta}({\bf P})=\Delta_{{\rm 2p}}({\bf P})$. As shown in
the main text, we assume that the MFR is due to the coupling between
the open channel $O$ and the bound state $|\phi_{\alpha}\rangle$
in the closed channel $C$, and the state $|\phi_{\beta}\rangle$
is far-off resonant to channel $O$. Thus, the direct coupling between
$|\phi_{\beta}\rangle$ and $O$ can be neglected. Explicitly, we
can make the approximation $\langle\phi_{\beta}|H_{{\rm rel}}({\bf P})|O\rangle_{I}=0$ which yields $\langle\phi_{\beta}|Y|O\rangle_{I}=0$. Substituting
Eq. (\ref{e1}) into (\ref{e2}), we obtain the linear equations for
$b_{{\bf k}}^{(\alpha,e,\beta)}$ which gives
\begin{align}
b_{{\bf k}}^{(l)} & =\sum_{l^{\prime}=\alpha,e,\beta}\{\left[E{\cal I}-\Sigma(E_{{\bf k}},{\bf P})\right]^{-1}\}_{l,l^{\prime}}\langle\phi_{l^{\prime}}|Y|\psi_{{\bf k}}^{{\rm bg(+)}}\rangle_{r}|O\rangle_{I}.\tag{S22}\label{bl}
\end{align}
Here $\{\left[E{\cal I}-\Sigma(E_{{\bf k}},{\bf P})\right]^{-1}\}_{l,l^{\prime}}$
denotes the $(l,l^{\prime})$-th element of the inverse matrix of
the $3\times3$ matrix $E{\cal I}-\Sigma(E_{{\bf k}},{\bf P})$, where
${\cal I}$ is the $3\times3$ identical matrix and $\Sigma(E,{\bf P})$
is the self-energy matrix which can be expressed as (in the basis
$\alpha,e,\beta$)
\begin{align}
\Sigma(E,{\bf P}) & =\left[\begin{array}{ccc}
E_{\alpha}+\langle\phi_{\alpha}|V_{{\rm hf}}G_{{\rm bg}}(E)V_{{\rm hf}}|\phi_{\alpha}\rangle & \Omega_{\alpha}^{\ast} & 0\\
\Omega_{\alpha} & \Delta_{{\rm 1p}}({\bf P})+\langle\phi_{e}|V_{FD}G_{D}(E)V_{FD}|\phi_{e}\rangle & \Omega_{\beta}^{\ast}\\
0 & \Omega_{\beta} & \Delta_{{\rm 2p}}({\bf P})
\end{array}\right],\tag{S23}\label{lk}
\end{align}
with
\begin{align*}
V_{{\rm hf}} & =V_{{\rm hf}}(r)|O\rangle_{I}\langle C|+h.c.;
\\ V_{FD}&=V_{FD}(r)|F\rangle_{I}\langle D|+h.c.;\\
G_{{\rm bg}}(E) & =\frac{1}{E+i0^{+}-\left(\frac{{\bf p}^{2}}{2\mu}+V_{{\rm bg}}(r)\right)};\\ G_{D}(E)&=\frac{1}{E+i0^{+}-\left(\frac{{\bf p}^{2}}{2\mu}+V_{D}(r)\right)}.
\end{align*}
Substituting Eq. (\ref{bl}) into Eqs. (\ref{e1}, \ref{ss}), we
can obtain the expression for $|\Phi_{{\bf k}}\rangle$ and the scattering
state $|\Psi_{{\bf k}}^{(+)}\rangle$.

\subsection{Low-Energy Expression for $\Sigma(E,{\bf P})$}

In this paper we consider the cases where $|E|$ is much smaller than
the characteristic energy of $V_{{\rm bg}}(r)$, i.e., the van der
Waals energy $E_{{\rm vdW}}$. For these cases, the expression (\ref{lk})
of $\Sigma(E,{\bf P})$ can be significantly simplified.


First, for real $E$, the factor $\langle\phi_{\alpha}|V_{{\rm hf}}G_{{\rm bg}}(E)V_{{\rm hf}}|\phi_{\alpha}\rangle$
can be re-expressed as

\begin{align}
\langle\phi_{\alpha}|V_{{\rm hf}}G_{{\rm bg}}(E)V_{{\rm hf}}|\phi_{\alpha}\rangle & =\int d{\bf q}\frac{|\langle\phi_{\alpha}|V_{{\rm hf}}|\psi_{{\bf q}}^{{\rm bg}(+)}\rangle_{r}|O\rangle_{I}|^{2}}{E+i0^{+}-E_{{\bf q}}}\nonumber \\
 & =\int d{\bf q}\frac{|\langle\phi_{\alpha}|V_{{\rm hf}}|\psi_{{\bf q}}^{{\rm bg}(+)}\rangle_{r}|O\rangle_{I}|^{2}}{-E_{{\bf q}}}+\int d{\bf q}|\langle\phi_{\alpha}|V_{{\rm hf}}|\psi_{{\bf q}}^{{\rm bg}(+)}\rangle_{r}|O\rangle_{I}|^{2}\left(\frac{1}{E+i0^{+}-E_{{\bf q}}}-\frac{1}{-E_{{\bf q}}}\right),\nonumber \\
\tag{S24}\label{e1-1}
\end{align}
In the region where $E_{{\bf q}}\gg|E|$, we have $\left|\frac{1}{E-E_{{\bf q}}}-\frac{1}{-E_{{\bf q}}}\right|\ll\left|\frac{1}{-E_{{\bf q}}}\right|.$
Thus, in this region the contribution of the to-be-integrated function
in the second term of the r.h.s. of Eq. (\ref{e1-1}) can be neglected.
Therefore, we can only do the second integration in the r.h.s. of
Eq. (\ref{e1-1}) in the region where $E_{{\bf q}}$ is not much larger
than $E$. On the other hand, in realistic ultracold atom systems,
the inter-channel coupling $V_{{\rm hf}}(r)$ is negligible in the
region $r\gtrsim r_{{\rm vdW}}$ with $r_{{\rm vdW}}$ being the van
der Waals radius. Thus, the the factor $|\langle\phi_{\alpha}|V_{{\rm hf}}|\psi_{{\bf q}}^{{\rm bg}(+)}\rangle_{r}|O\rangle_{I}|^{2}$
is determined by the behavior of the wave function $_{r}\!\langle{\bf r}|\psi_{{\bf q}}^{{\rm bg}(+)}\rangle_{r}$
in the region $r\lesssim r_{{\rm vdW}}$, where $V_{{\rm bg}}(r)$
is a deep potential well and thus $_{r}\!\langle{\bf r}|\psi_{{\bf q}}^{{\rm bg}(+)}\rangle_{r}$
is almost independent of ${\bf q}$ for $E_{{\bf q}}<E_{{\rm vdW}}$.
Therefore, in our calculation we make the approximation
\begin{align*}
\int d{\bf q}|\langle\phi_{\alpha}|V_{{\rm hf}}|\psi_{{\bf q}}^{{\rm bg}(+)}\rangle_{r}|O\rangle_{I}|^{2}\left(\frac{1}{E+i0^{+}-E_{{\bf q}}}-\frac{1}{-E_{{\bf q}}}\right) & \approx|\langle\phi_{\alpha}|V_{{\rm hf}}|\psi_{{\bf q}=0}^{{\rm bg}(+)}\rangle_{r}|O\rangle_{I}|^{2}\int d{\bf q}\left(\frac{1}{E+i0^{+}-E_{{\bf q}}}-\frac{1}{-E_{{\bf q}}}\right)\\
 & =\eta(E)|\langle\phi_{\alpha}|V_{{\rm hf}}|\psi_{{\bf q}=0}^{{\rm bg}(+)}\rangle_{r}|O\rangle_{I}|^{2}4\pi^{2}\mu\sqrt{2\mu},\tag{S25}
\end{align*}
where
\[
\eta(E)=\left\{ \begin{array}{cc}
-i\sqrt{E}, & {\rm for}\ E>0\\
\sqrt{(-E)}, & {\rm for}\ E<0\ {\rm or}\ {\rm Im}[E]\neq0
\end{array}\right.,\tag{S26}
\]
with $\sqrt{z}=\sqrt{|z|}e^{i\arg[z]/2}$ and $\arg[z]\in(-\pi,+\pi]$.
Thus, we have
\begin{equation}
\langle\phi_{\alpha}|V_{{\rm hf}}G_{{\rm bg}}(E)V_{{\rm hf}}|\phi_{\alpha}\rangle\approx E_{\alpha}^{\prime}+\chi(E),\tag{S27}\label{app1}
\end{equation}
where
\begin{align}
E_{\alpha}^{\prime} & =\int d{\bf q}\frac{|\langle\phi_{\alpha}|V_{{\rm hf}}|\psi_{{\bf q}}^{{\rm bg}(+)}\rangle_{r}|O\rangle_{I}|^{2}}{-E_{{\bf q}}};\tag{S28}\label{ealphap}\\
\chi(E) & =\eta(E)|\langle\phi_{\alpha}|V_{{\rm hf}}|\psi_{{\bf q}=0}^{{\rm bg}(+)}\rangle_{r}|O\rangle_{I}|^{2}4\pi^{2}\mu\sqrt{2\mu}.\tag{S29}\label{jianwen}
\end{align}

Second, as shown above, in our system $E_{D}-E_e$ is much larger then both $\Delta_{\rm 1p}({\bf P})$ and $E$.
As a result, the factor $\langle\phi_{e}|V_{FD}G_{D}(E)V_{FD}|\phi_{e}\rangle$ is almost independent of $E$, and satisfies \cite{paulOpticalFRs}
\begin{equation}
{\rm Im}\left[\langle\phi_{e}|V_{FD}G_{D}(E)V_{FD}|\phi_{e}\rangle\right]=-i\frac{\gamma}{2},\tag{S30}\label{dec-1}
\end{equation}
with $\gamma$ being the decay rate of $|\phi_{e}\rangle$. In addition,
${\rm Re}\left[\langle\phi_{e}|V_{FD}G_{D}(E)V_{FD}|\phi_{e}\rangle\right]$ is the Lamb shift of the energy of $|\phi_{e}\rangle$ , which
is induced by the coupling between $|\phi_{e}\rangle$ and the channel
$D$. It can be absorbed in the definition of $E_{e}$. Explicitly, we can make
the notation replacement
\begin{equation}
E_{e}+{\rm Re}\left[\langle\phi_{e}|V_{FD}G_{D}(E)V_{FD}|\phi_{e}\rangle\right]\rightarrow E_{e}.\tag{S31}\label{nr-1}
\end{equation}
Substituting Eqs. (\ref{app1},\ref{dec-1},\ref{nr-1}) into Eq.
(\ref{lk}), we obtain
\begin{equation}
\Sigma(E,{\bf P})\approx\left[\begin{array}{ccc}
E_{\alpha}+E_{\alpha}^{\prime}+\chi(E) & \Omega_{\alpha}^{\ast} & 0\\
\Omega_{\alpha} & \Delta_{{\rm 1p}}({\bf P})-i\frac{\gamma}{2} & \Omega_{\beta}^{\ast}\\
0 & \Omega_{\beta} & \Delta_{{\rm 2p}}({\bf P})
\end{array}\right],\tag{S32}\label{lkapp}
\end{equation}
Our following calculation are based on this low-energy expression of $\Sigma(E,{\bf P})$.

\subsection{3D Scattering Amplitude}

Now we consider the elastic scattering amplitude in the open channel
$O$. According to the formal scattering theory, for our system this
scattering amplitude can be expressed as
\begin{align}
f & =-(2\pi)^{2}\mu\left[_{I}\langle O|_{r}\langle{\bf k}^{\prime}|(V+Y+Z)|\Psi_{{\bf k}}^{(+)}\rangle\right],\tag{S33}\label{sf}
\end{align}
where ${\bf k}$ and ${\bf k}^{\prime}$ are the incident and output
momentum, respectively, and satisfies $|{\bf k}|=|{\bf k}^{\prime}|=k$,
and the operator $V$ is defined as
\[
V=V_{{\rm bg}}(r)|O\rangle_{I}\langle O|+V_{D}(r)|D\rangle_{I}\langle D|.\tag{S34}
\]
Substituting the expression of the scattering state $|\Psi_{{\bf k}}^{(+)}\rangle$
derived in Sec. II.B into Eq. (\ref{sf}) and doing some
straightforward calculations, we obtain
\begin{align}
f & =f^{{\rm (bg)}}-(2\pi)^{2}\mu\sum_{l=\alpha,e,\beta}\ _{I}\langle O|_{r}\langle\psi_{{\bf k}^{\prime}}^{{\rm bg(-)}}|Y|\phi_{l}\rangle b_{{\bf k}}^{(l)},\tag{S35}\label{ff}
\end{align}
with $b_{{\bf k}}^{(l)}$ being given by Eq. (\ref{bl}). Here $f^{{\rm (bg)}}=-(2\pi)^{2}\mu\left[|_{r}\langle{\bf k}^{\prime}|V_{{\rm bg}}(r)|\psi_{{\bf k}}^{{\rm bg(+)}}\rangle\right]$
is the background scattering amplitude for the open channel itself,
and $|\psi_{{\bf k}^{\prime}}^{{\rm bg(-)}}\rangle$ is the in-coming
scattering state for the open channel itself, which is defined as
\begin{align}
|\psi_{{\bf k}^{\prime}}^{{\rm bg(-)}}\rangle_{r} & =\lim_{\varepsilon\rightarrow0^{-}}\frac{i\varepsilon}{E_{{\bf k}^{\prime}}+i\varepsilon-\left(\frac{{\bf p}^{2}}{2\mu}+V_{{\rm bg}}(r)\right)}|{{\bf k}^{\prime}}\rangle_{r}\tag{S36}\label{fff}
\end{align}
and satisfies
\begin{align}
_{r}\langle\psi_{{\bf k}^{\prime}}^{{\rm bg(-)}}| & =\ _{r}\!\langle{\bf k}^{\prime}|+\ _{r}\!\langle{\bf k}^{\prime}|V_{{\rm bg}}(r)G_{\rm bg}(E_{{\bf k}^\prime}).\tag{S37}
\end{align}

Substituting Eq. (\ref{bl}) into Eq. (\ref{ff}), we can further
derive
\begin{equation}
f=f^{{\rm (bg)}}-(2\pi)^{2}\mu e^{2i\delta_{E_{{\bf k}}}^{({\rm bg})}}|\langle\phi_{\alpha}|V_{{\rm hf}}|\psi_{{\bf k}}^{{\rm bg}(+)}\rangle_{r}|O\rangle_{I}|^{2}\{\left[E_{{\bf k}}{\cal I}-\Sigma(E_{{\bf k}},{\bf P})\right]^{-1}\}_{\alpha\alpha}.\tag{S38}\label{f2}
\end{equation}
Here $\delta_{E_{{\bf k}}}^{({\rm bg})}$ is the $s$-wave phase shift
for the open channel $O$ itself and we have used the relation $_{r}\langle\psi_{{\bf k}}^{{\rm bg}(-)}|{\bf r}\rangle_{r}=e^{2i\delta_{E_{{\bf k}}}^{({\rm bg})}}{}_{r}\langle\psi_{{\bf k}}^{{\rm bg}(+)}|{\bf r}\rangle_{r}$,
with $|{\bf r}\rangle_{r}$ being the eigen-state of the two-atom
relative position operator. This relation can be proved via the Schr$\ddot{{\rm o}}$dinger
equation in the spatial space as well as the out-going and in-coming
boundary conditions satisfied by $_{r}\langle{\bf r}|\psi_{{\bf k}}^{{\rm bg}(\pm)}\rangle_{r}$.
Furthermore, as shown above, in our problem the factor $|\langle\phi_{\alpha}|V_{{\rm hf}}|\psi_{{\bf k}}^{{\rm bg}(+)}\rangle_{r}|O\rangle_{I}|^{2}$
changes very slowly with ${\bf k}$ for $E_{{\bf k}}\ll E_{{\rm vdW}}$.
Therefore, we can make the approximation
\[
|\langle\phi_{\alpha}|V_{{\rm hf}}|\psi_{{\bf k}}^{{\rm bg}(+)}\rangle_{r}|O\rangle_{I}|^{2}\approx|\langle\phi_{\alpha}|V_{{\rm hf}}|\psi_{{\bf k}=0}^{{\rm bg}(+)}\rangle_{r}|O\rangle_{I}|^{2}\tag{S39}
\]
in the r.h.s. of Eq. (\ref{f2}), and obtain
\begin{equation}
f=f^{{\rm (bg)}}-(2\pi)^{2}\mu e^{2i\delta_{E_{{\bf k}}}^{({\rm bg})}}|\langle\phi_{\alpha}|V_{{\rm hf}}|\psi_{{\bf k}=0}^{{\rm bg}(+)}\rangle_{r}|O\rangle_{I}|^{2}\{\left[E_{{\bf k}}{\cal I}-\Sigma(E_{{\bf k}},{\bf P})\right]^{-1}\}_{\alpha\alpha},\tag{S40}\label{f2-1}
\end{equation}
with $\Sigma(E_{{\bf k}},{\bf P})$ given by Eq. (\ref{lkapp}). In the following we use the result (\ref{f2-1}) to calculate the
scattering length.

\subsection{Proof of Eq. (8)}

Now we calculate the scattering length $a$ for zero scattering energy
and proof Eq. (8) of the main text. The scattering length is defined
as
\begin{align}
a=-f|_{{\bf k}={\bf k}^{\prime}=0}.\tag{S41}\label{ad}
\end{align}
We first consider the case without Raman beams (i.e., $Z=0$). Taking
${\bf k}={\bf k}^{\prime}=0$ for Eq. (\ref{f2-1}) and using the
fact $\delta_{E_{{\bf k=0}}}^{({\rm bg})}=0$, we obtain
\begin{align}
a=a_{{\rm bg}}-(2\pi)^{2}\mu\frac{|\langle\phi_{\alpha}|V_{{\rm hf}}|\psi_{{\bf k}=0}^{{\rm bg}(+)}\rangle_{r}|O\rangle_{I}|^{2}}{E_{\alpha}+E_{\alpha}^{\prime}},\tag{S42}
\end{align}
where $a_{{\rm bg}}=-f^{{\rm (bg)}}|_{{\bf k}={\bf k}^{\prime}=0}$
is the background scattering length, and $E_{\alpha}^{\prime}$ is
defined in Eq. (\ref{ealphap}). Comparing this result with the usual
expression for the scattering length near a MFR
\[
a=a_{{\rm bg}}\left(1-\frac{\Delta_{B}}{B-B_{0}}\right),\tag{S43}
\]
we obtain the relations
\begin{align}
(2\pi)^{2}\mu|\langle\phi_{\alpha}|V_{{\rm hf}}|\psi_{{\bf k}=0}^{{\rm bg}(+)}\rangle_{r}|O\rangle_{I}|^{2} & =(\delta\mu)a_{{\rm bg}}\Delta_{B};\tag{S44a}\label{r1}\\
E_{\alpha}+E_{\alpha}^{\prime} & =(\delta\mu)(B-B_{0}),\tag{S44b}\label{r2}
\end{align}
where $\delta\mu$ is the magnetic momentum difference between the
open channel $O$ and the closed channel $C$, as defined in the main
text.

Now we consider the case with Raman beams. In this case we can also
calculate the scattering length $a$ by taking ${\bf k}={\bf k}^{\prime}=0$
for Eq. (\ref{f2-1}). Using Eqs. (\ref{r1}, \ref{r2}, \ref{dec-1},
\ref{lkapp}), we finally obtain
\[
a({\bf P})\!=\!a_{{\rm bg}}-\frac{(\delta\mu)a_{{\rm bg}}\Delta_{B}\left[\Delta_{{\rm 1p}}({\bf P})-i\frac{\gamma}{2}-\frac{|\Omega_{\beta}|^{2}}{\Delta_{{\rm 2p}}({\bf P})}\right]}{(\delta\mu)(B-B_{0})\!\left[\Delta_{{\rm 1p}}({\bf P})-i\frac{\gamma}{2}-\frac{|\Omega_{\beta}|^{2}}{\Delta_{{\rm 2p}}({\bf P})}\right]\!-|\Omega_{\alpha}|^{2}}.\ \tag{S45} \label{aaa}
\]
That is Eq. (8) in the main text.

\section{3D Bound-State Energy}

Now we calculate the energy $E_{b}$ of the 2-body bound state $|\Phi_{b}\rangle$.
Here we consider the case with $|E_{b}|\ll E_{{\rm vdW}}$. Since
we have assumed that the threshold energy of the the auxiliary channel
$D$ is much lower than $-E_{{\rm vdW}}$, precisely speaking there
is no bound state with $|E_{b}|\ll E_{{\rm vdW}}.$ Nevertheless,
our system may have approximate bound state (quasi bound state) with
complex energy $E_{b}$ which satisfies
\[
{\rm Re}[E_{b}]<0,\ \ |{\rm Re}[E_{b}]|\ll E_{{\rm vdW}},\ \ {\rm Im}[E_{b}]<0.
\]
Here ${\rm Re}[E_{b}]$ and $|{\rm Im}[E_{b}]|$ describe the energy
and decay rate of this approximate bound state, respectively. Similar
as above, for our system both $|\Phi_{b}\rangle$ and $E_{b}$ are
functions of the of the CoM momentum ${\bf P}$.

We can obtain the approximate bound state via the equation
\begin{equation}
H_{{\rm eff}}({\bf P})|\Phi_{b}\rangle=E_{b}({\bf P})|\Phi_{b}\rangle.\tag{S46}\label{heffse}
\end{equation}
Here $H_{{\rm eff}}({\bf P})$ is the effective Hamiltonian for our
system and can be expressed as
\begin{equation}
H_{{\rm eff}}({\bf P})=H_{F}({\bf P})+V_{{\rm hf}}+Z-i\frac{\gamma}{2}|\phi_{e}\rangle\langle\phi_{e}|.\tag{S47}\label{heffp}
\end{equation}
We can further express $|\Phi_{b}\rangle$ as $|\Phi_{b}\rangle=|\phi_{O}\rangle_{r}|O\rangle_{I}+\sum_{l=\alpha,e,\beta}c_{l}|\phi_{l}\rangle$.
Substituting this expression into Eq. (\ref{heffse}), we can derive
the equations for $|\phi_{O}\rangle_{r}$ and the coefficients $c_{l}$
($l=\alpha,e,\beta$). Eliminating $|\phi_{O}\rangle_{r}$ from these
equations and using Eq. (\ref{app1},\ref{ealphap},\ref{jianwen}),
we obtain
\begin{equation}
\left[E_{b}{\cal I}-\Sigma(E_{b},{\bf P})\right]\left[\begin{array}{c}
c_{\alpha}\\
c_{e}\\
c_{\beta}
\end{array}\right]=0.\tag{S48}\label{ebeq}
\end{equation}
with $\Sigma(E_{b},{\bf P})$ being given by Eq. (\ref{lkapp}). Eq.
(\ref{ebeq}) yields that the energy $E_{b}$ is determined by the
equation
\begin{equation}
\det\left[E_{b}{\cal I}-\Sigma(E_{b},{\bf P})\right]=0.\tag{S49}\label{ebeq2}
\end{equation}
We derive $E_{b}$ by numerically solving Eq. (\ref{ebeq2}).

\section{Robustness of our results}

Now we consider the robustness of our result with respect to the uncertainty of the MFR point $B_0$ and width $\Delta_B$. We first study the effect induced by the uncertainty of $B_0$  for the ultracold $^{40}$K gas discussed in our main text. For this system the
main value of $B_0$ is 202.2G and the uncertainty of $B_0$ is 0.02G \cite{jingbb}.
Below in Fig.~S2(a) we show the behavior of ${\rm Re}[a(P_x)]$ 
for the cases where $B_0=202.2$G and $202.2\pm 0.02$G while $B$ is fixed at $202.2{\rm G}-0.07\Delta_B=201.64$G (corresponding to the case in Fig.~2(a) of the main text). Similarly, in Fig.~S2(b) and Fig.~S2(c) we show the results for $B=202.2{\rm G}+0.07\Delta_B=202.76$G (corresponding to the case in Fig.~2(b) of the main text) and $B=202.2{\rm G}$, respectively. These figures shows that our results are robust with respect to the uncertainty of the MFR point.

\begin{figure*}[htb]
\includegraphics[width=17cm]{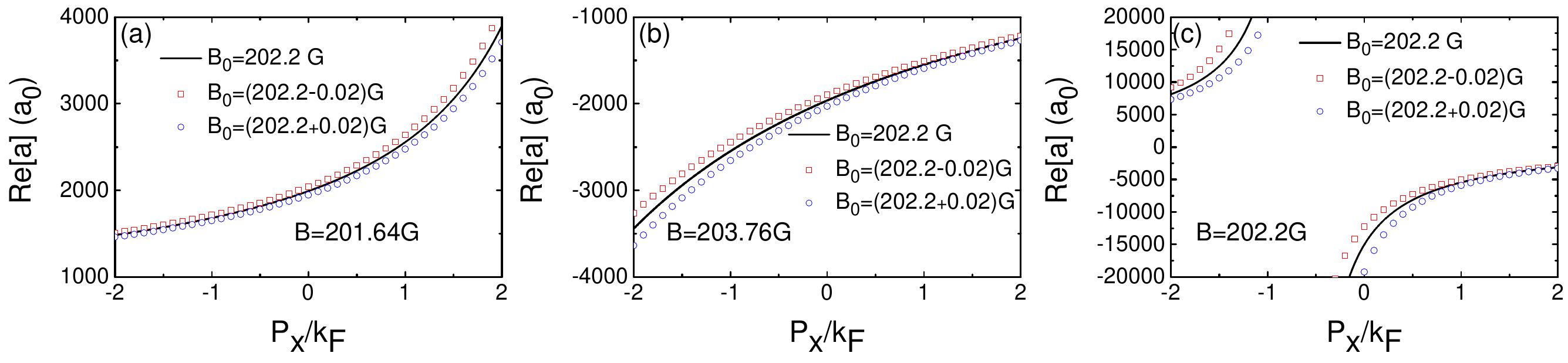}
{Figure S2: (color online) ${\rm Re}[a(P_x)]$ of ultracold $^{40}$K
gases with MFR point $B_0=$202.2G (black solid line),  $B_0$=(202.2-0.02)G (red squares) and   $B_0$=(202.2+0.02)G (blue circles). Here we consider the cases with fixed value of magnetic field $B$. {\bf (a)}: $B$=202.2G-0.07$\Delta_B$=201.64G (corresponding to the case in Fig.~2(a) of the main text), {\bf (b)}: $B=202.2{\rm G}+0.07\Delta_B=202.76$G (corresponding to the case in Fig.~2(b) of the main text) and {\bf (c)}: $B=202.2$G. Other parameters of Fig.~(a) and Fig.~(b, c) are same as the ones in Fig.~2(a) and Fig.~2(b) of the main text, respectively.
}
\end{figure*}

We can also understand our above conclusion with Fig.~S3, where we illustrate variation of the scattering length $a(P_x)$ with the magnetic field $B$, for the case in Fig.~2(b) of our main text. It is shown that for different CoM momentum $P_x$ the scattering length has different resonant point in the $B$-axis. Therefore, loosely speaking, in our system the Doppler effect induces a {\it CoM momentum dependent} shift of the MFR point, and make the scattering length to be CoM momentum dependent. Fig.~S3 clearly show that, when $P_x$ is modified from $-2k_F$ to $2k_F$ the MFR point is shifted by about $0.6$G, which is much larger than the uncertainty of  the MFR point ($0.02$G). Therefore,  our results are quite robust for that uncertainty.
\begin{figure*}[htp]
\includegraphics[width=8cm]{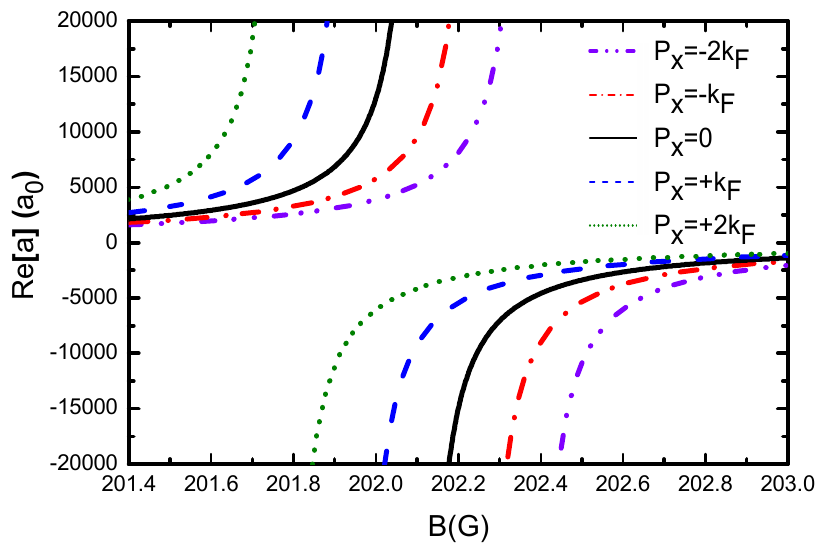}
\begin{center}
{FIG. S3: (color online) The scattering length $a(P_x)$ as a function of $B$, for $P_x=0$ (black solid line), $P_x=-2k_F$ (violet dashed-doted-dotted line), $P_x=-k_F$ (red dashed-dotted line),$P_x=+k_F$ (blue dashed line) and $P_x=+2k_F$ (green dotted line). The parameters in this figure are same as in Fig.~2(b) of our main text.}
\end{center}
\end{figure*}

Now we investigate the effect of the uncertainty of the resonance width $\Delta_B$. To this end we consider two cases with resonance width $\Delta_B$ and $\Delta_B+\delta_B$, where $\delta_B$ is the uncertainty of the resonance width. Eq. (8) of our main text clearly show that the scattering lengths for these two cases satisfy the relation
\begin{equation}
\frac{a({\bf P},\Delta_B+\delta_B)-a_{\rm bg}}{a({\bf P},\Delta_B)-a_{\rm bg}}=\frac{\Delta_B+\delta_B}{\Delta_B}\tag{S50}.
\end{equation}
When the magnetic field is close to the resonance point, e.g., in the cases in Figs.~2(a, b) in the main text, we have $|a({\bf P},\Delta_B)|\gg |a_{\rm bg}|$ and $|a({\bf P},\Delta_B+\delta_B)|\gg |a_{\rm bg}|$, which yields
\begin{equation}
\frac{a({\bf P},\Delta_B+\delta_B)-a({\bf P},\Delta_B)}{a({\bf P},\Delta_B)}=\frac{\delta_B}{\Delta_B}\tag{S51},
\end{equation}
i.e., the relative error of the scattering length is just the relative uncertainty $\delta_B/\Delta_B$ of the resonance width. In realistic systems we usually have $|\delta_B/\Delta_B|\ll 1$. Nevertheless, the Doppler-effect-induced relative variation of scattering length with cener-of-mass momentum,(for our example of $^{40}$K atoms it is $\frac{|a(P_x=2k_F)-a(P_x=-2k_F)|}{|a(P_x=2k_F)|}$) can be of the order of 1 or even larger, as shown in Figs~2(a, b). Thus, the Doppler effect is robust with respect to the uncertainty of the resonance width.

\section{Quasi-1D system}

In this subsection we consider an two-component Fermi gas in an quasi-1D
confinement along the $x$-direction, as described in the main text.
In this system the effective low-energy 1D interaction between two
atoms in different components can be expressed as
\begin{equation}
V_{{\rm 1D}}=g_{{\rm 1D}}\delta(x)\equiv-\frac{1}{\mu a_{{\rm 1D}}}\delta(x),\tag{S52}\label{g1d}
\end{equation}
where the 1D scattering length $a_{{\rm 1D}}$
is given by \cite{olshaniis}
\begin{equation}
a_{{\rm 1D}}=-\frac{b_{\perp}^{2}}{2a^{(\omega_{\perp})}}\left(1-{\cal C}\frac{a^{(\omega_{\perp})}}{b_{\perp}}\right).\tag{S53}\label{a1d}
\end{equation}
Here $b_{\perp}=\sqrt{1/(\mu\omega_{\perp})}$ is the characteristic
length of the transverse trap and ${\cal C}=-\zeta(1/2)=1.4603...$.
In Eq. (\ref{a1d}) $a^{(E)}$ is the ``energy-dependent\char`\"{}
3D $s$-wave scattering length of these two atoms, which is defined
as
\begin{equation}
a^{(E)}=\frac{-1}{\cot\delta_{E}\sqrt{2\mu E}}\tag{S54}\label{ae-1}
\end{equation}
with $\delta_{E}$ being the $s$-wave phase shift corresponding to
scattering energy $E$. It relates to the scattering length $a$ we
studied before via the relation $a^{(E=0)}=a$. When the two-atom interaction is controlled by a MFR modulated
by two Raman beams, the scattering length $a^{(E)}$ becomes a function
of the CoM momentum ${\bf P}$. This scattering length can be calculated
as follows. Using Eq. (\ref{f2-1}) and the relation $f^{({\rm bg})}=-1/[ik-k\cot\delta_{E_{{\bf k}}}^{({\rm bg})}]$,
we can re-express Eq. (\ref{f2-1}) as
\begin{equation}
f=\frac{-1}{ik-k\cot\delta_{E_{{\bf k}}}^{({\rm bg})}}-\frac{e^{2i\delta_{E_{{\bf k}}}^{({\rm bg})}}}{ik+A_{{\bf k}}({\bf P})}\tag{S55}\label{fnew}
\end{equation}
where
\begin{equation}
A_{{\bf k}}({\bf P})=\frac{1}{(2\pi)^{2}\mu|\langle\phi_{\alpha}|W|\psi_{{\bf k}=0}^{{\rm bg}(+)}\rangle_{r}|O\rangle_{I}|^{2}\{\left[E_{{\bf k}}{\cal I}-\Sigma(E_{{\bf k}},{\bf P})\right]^{-1}\}_{\alpha\alpha}}-ik.\tag{S56}\label{akp}
\end{equation}
with $\Sigma(E_{{\bf k}},{\bf P})$ given by Eq. (\ref{lkapp}). On
the other hand, Eq. (\ref{ae}) yields $f=-1/[ik+1/a^{(E_{{\bf k}})}]$.
Comparing this result and Eq. (\ref{fnew}) and using the relation
$-k\cot\delta_{E_{{\bf k}}}^{({\rm bg})}=1/a_{{\rm bg}}$ , we obtain
the expression of $a^{(E_{{\bf k}})}$:
\begin{equation}
a^{(E_{{\bf k}})}=\frac{a_{{\rm bg}}A_{{\bf k}}({\bf P})+1}{A_{{\bf k}}({\bf P})-k^{2}a_{{\rm bg}}}.\tag{S57}\label{ae}
\end{equation}
Here we consider the case in which the Raman beams are counterpropagating
along the $x$-axis. It is clear that in this case $a^{(E_{{\bf k}})}$
only depends on the value of $E_{{\bf k}}$ and the $x$-component
of ${\bf P}$ (i.e., $P_{x}$), and are independent of the direction
of ${\bf k}$ and the values of $P_{y}$ and $P_{z}$.

\begin{figure*}[t]
\includegraphics[width=7.5cm]{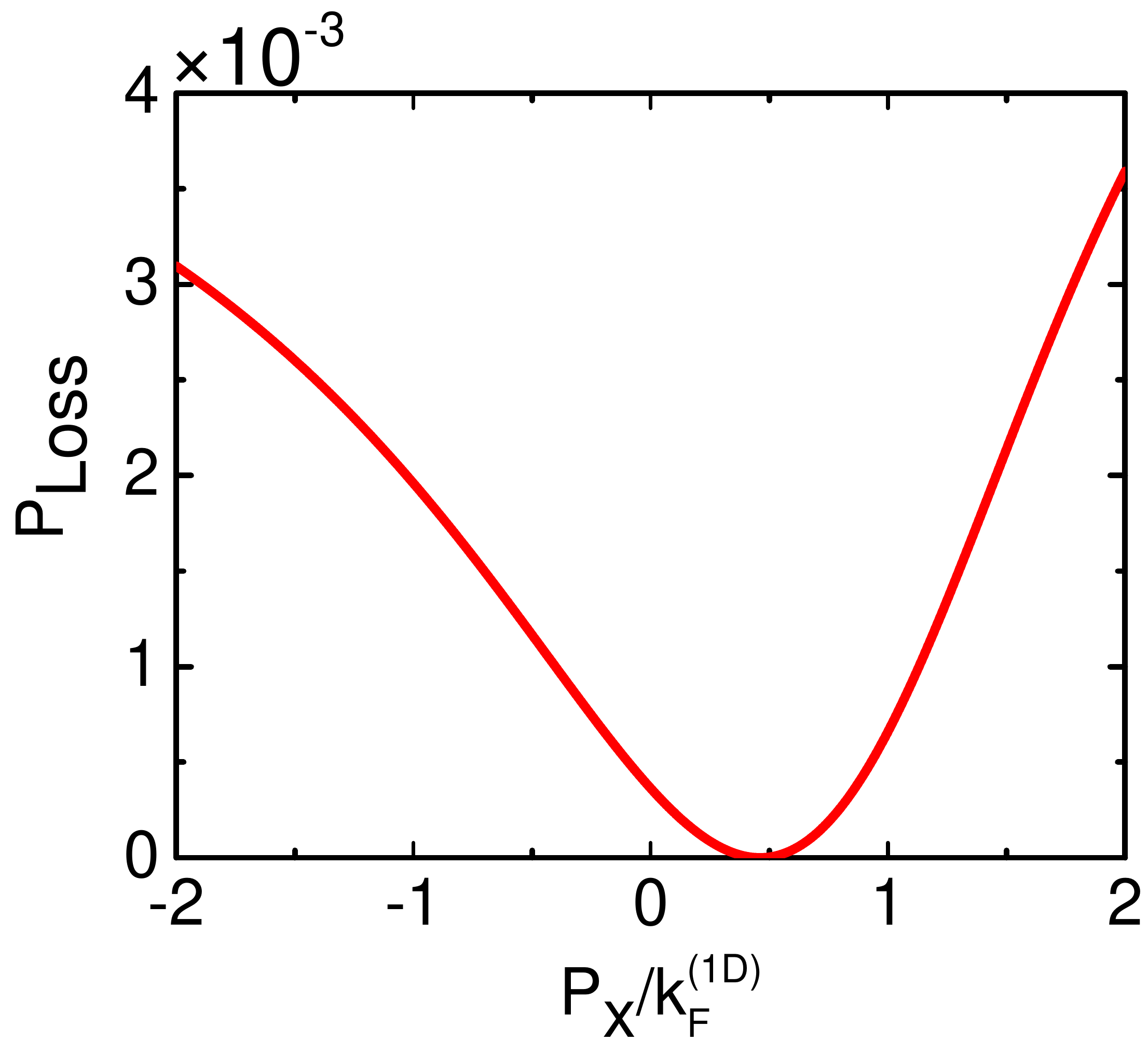}
\begin{center}
{FIG. S4: (color online) The loss probability $P_{{\rm loss}}$ for the quasi-1D
system studied in Fig.~3 of the main text. }
\end{center}
\end{figure*}

Substituting Eq. (\ref{ae}) into Eqs. (\ref{a1d}, \ref{g1d}), we
can obtain the expressions of the effective 1D scattering length $a_{{\rm 1D}}$
and the the effective 1D interaction intensity $g_{{\rm 1D}}$, which
are both $P_{x}$-dependent. In particular, according to Eqs. (\ref{g1d})
and (\ref{a1d}), when
 $a^{(\omega_{\perp})}(P_{x})$ is tuned to satisfy $a^{(\omega_{\perp})}(P_{x}=0)=a_{\perp}/{\cal C}$
we can obtain an interesting quasi-1D system with
$g_{{\rm 1D}}(P_{x}=0)=\infty$, $g_{{\rm 1D}}(P_{x}>0)<0$ and $g_{{\rm 1D}}(P_{x}<0)>0$,
as shown in Fig.~3 of the main text.

Finally, we consider the two-body collisional loss of the quasi-1D
gas of two-component $^{40}$K atoms studied in Fig.~3 of the main text,
 where the one-body momentum $k_x$ along the $x$-direction is in the region between $\pm k_F^{\rm (1D)}\equiv\pm3/(4b_{\perp})$ and thus we have $P_x\in[-2 k_F^{\rm (1D)}, 2k_F^{\rm (1D)}]$. For one quasi-1D scattering process, the two-body
collisional loss probability $P_{{\rm loss}}$ is given by $P_{{\rm loss}}=1-|r|^{2}-|t|^{2}$,
where $r=-1/(1+ik_{x}a_{{\rm 1D}})$ and $t=1+r$ are the reflection
and transmission amplitude of this scattering process, respectively,
with $k_{x}$ being the incident momentum along the $x$-direction.
For our system $P_{{\rm loss}}$ also depends on the CoM momentum
$P_{x}$. In Fig.~S4 we illustrate $P_{{\rm loss}}$ for our system with $k_{x}=k_{F}^{{\rm (1D)}}$.
It is shown that for this system the loss probability is at most on the order of
$10^{-3}$.

With this result, we can further estimate the life time of our system. First, the density $n_{{\rm 1D}}$ of a quasi-1D ultracold gas
is related to the 1D Fermi momentum $k_{F}^{{\rm (1D)}}$ via $n_{{\rm 1D}}=k_{F}^{{\rm (1D)}}/\pi$.
Thus, the average distance of two atoms is $d=1/n_{{\rm 1D}}=\pi/k_{F}^{{\rm (1D)}}$.
Second, the maximum relative velocity of two atoms is $v_{{\rm max}}=k_{F}^{{\rm (1D)}}/m$, with $m$ being the single-atom mass.
Thus, the frequency of the collision between a given atom and other
atoms is at most $\nu=v_{{\rm max}}/d$. Therefore, the life time
for each atom in the ultracold gas can be estimated as $\tau_{{\rm 1D}}=1/(\nu P_{{\rm loss}})=m\pi/(\hbar k_{F}^{{\rm (1D)2}}P_{{\rm loss}})$.
Substituting the parameters of Fig.~3 of the
main text into this formula, we obtain $\tau_{{\rm 1D}}>0.04$s.

\end{widetext}